\documentclass[letterpaper]{article} 
\usepackage{aaai25}  
\usepackage{times}  
\usepackage{helvet}  
\usepackage{courier}  
\usepackage[hyphens]{url}  
\usepackage{graphicx} 
\usepackage{array}         
\usepackage{booktabs}      
\usepackage{longtable}     
\usepackage{tabularx}      
\usepackage{multirow}      
\usepackage{caption}       
\usepackage{natbib}  
\usepackage{caption} 
\usepackage{algorithm}
\usepackage{algorithmic}

\usepackage{multirow}
\usepackage{amsmath}
\usepackage{xcolor}

\usepackage{newfloat}
\usepackage{listings}
\DeclareCaptionStyle{ruled}{labelfont=normalfont,labelsep=colon,strut=off} 
\floatstyle{ruled}
\newfloat{listing}{tb}{lst}{}
\floatname{listing}{Listing}
\title{AI vs. Human Paintings? Deciphering Public Interactions and Perceptions towards AI-Generated Paintings on TikTok}
\author{
    Jiajun Wang\equalcontrib\textsuperscript{\rm 1}\textsuperscript{\textsection},
    Xiangzhe Yuan\equalcontrib\textsuperscript{\rm 1}\textsuperscript{\textdaggerdbl},
    Siying Hu\textsuperscript{\rm 1},
    Zhicong Lu\textsuperscript{\textdagger}\textsuperscript{\rm 1}\textsuperscript{\#}
}
\affiliations{
    \textsuperscript{\rm 1}Department of Computer Science\\

    City University of Hong Kong\\
    Hong Kong SAR\\
jiajwang02@gmail.com, yuanxiangzhe9@gmail.com, siyinghu-c@my.cityu.edu.hk, zhiconlu@cityu.edu.hk
}

\begin{document}

\maketitle
{\renewcommand{\thefootnote}{\textdagger} 
\footnotetext{Corresponding author: Zhicong Lu, Email: zhiconlu@cityu.edu.hk.}}

{\renewcommand{\thefootnote}{\textsection} 
\footnotetext{Current affiliation: School of Information Systems and Technology, University of New South Wales,
Sydney, Australia. New Email: jiajun.wang5@student.unsw.edu.au}}

{\renewcommand{\thefootnote}{\textdaggerdbl} 
\footnotetext{Current affiliation: University of Iowa, Iowa, United States.}}

{\renewcommand{\thefootnote}{\#} 
\footnotetext{Current affiliation: George Mason University, Virginia, United States. New Email: zlu6@gmu.edu}}

\begin{abstract}
With the development of generative AI technology, a vast array of AI-generated paintings (AIGP) have gone viral on social media like TikTok. However, some negative news about AIGP has also emerged. For example, in 2022, numerous painters worldwide organized a large-scale anti-AI movement because of the infringement in generative AI model training. This event reflected a social issue that, with the development and application of generative AI, public feedback and feelings towards it may have been overlooked. Therefore, to investigate public interactions and perceptions towards AIGP on social media, we analyzed user engagement level and comment sentiment scores of AIGP using human painting videos as a baseline. In analyzing user engagement, we also considered the possible moderating effect of the aesthetic quality of Paintings. Utilizing topic modeling, we identified seven reasons, including hyperrealistic quality, ambivalent reactions, perceived theft of art, etc., leading to negative public perceptions of AIGP. Our work may provide instructive suggestions for future generative AI technology development and avoid potential crises in human-AI collaboration.
\end{abstract}

\section{Introduction}
With the rise of various Generative AI technologies such as ``Midjourney'', ``stable diffusion'' and ``ChatGPT'' in 2022, a vast array of AI-generated content (AIGC), such as images, texts, sounds, and even videos, has widely spread across major social media platforms. AIGC's powerful functions and ease of use have gained the favor of many industries, particularly the art industry~\cite{hitsuwari2023does}. Many famous artists, such as Sougwen Chung, Helena Sarin, Refik Anadol, etc., have created advanced and groundbreaking artworks in cooperation with generative AI. As Sougwen Chung said\footnote{https://zhuanlan.zhihu.com/p/149217204}, `` As an artist working with these AI-generated tools, the prospect of artificial intelligence offers me a fresh perspective.'' The AIGC has given the art industry a lot of new opportunities and provided artists with motivation.

However, the public has shown resistance towards genAI art. In December 2022, many painters uploaded images stating \textit{``NO TO AI-GENERATED IMAGES''\footnote{https://arstechnica.com/information-technology/2022/12/artstation-artists-stage-mass-protest-against-ai-generated-artwork/}} and called for a global boycott of AI-generated paintings (AIGP). This incident started because some artists felt that AI paintings had plagiarized their paintings without consent and that their copyrights had been violated. This struggle lasted for nearly a year, and until now, many painters are still calling for a ban on uploading AI-generated paintings on social media.

As AIGC circulates widely on social media, scholars have investigated how factors like text quality~\cite{zhang2023human} and AI usage~\cite{messer2024co} affect viewers' perceptions. However, many overlook AI's painting capabilities and the aesthetic quality of AIGP on viewer participation and attitudes. Ragot et al. ~\cite{ragot2020ai} found significant cognitive bias in public perceptions of AIGP, while Latikka et al.'s study ~\cite{latikka2023ai} found that the public's attitude toward AI in the field of art was generally negative. However, their research did not consider Image Aesthetic Quality (IAQ) as a variable. Other studies~\cite{bhandari2019understanding, shi2021effects} have shown that aesthetic quality significantly influences decision-making and perceptions. Therefore, we included IAQ as a potential moderating variable affecting user engagement.

This paper presented a comprehensive analysis of public perceptions towards AIGP based on user engagement, comment sentiment scores, and topic modeling. User engagement is a multidimensional concept that encompasses not just behavioral aspects (actions) but also emotional aspects (feelings)~\cite{khan2017social,hollebeek2011exploring}. However, high levels of engagement can only show users' interaction intention while cannot indicate completely positive perceptions. Considering this, we also adopted comment sentiment analysis to compare the emotional scores of comments between AIGP and human paintings for further perception analysis.  Based on the above research motivations and feasibility, we formulated our research questions as follows:

\vspace{12pt}
\textbf{RQ1: }How are people's engagement levels with AIGP on social media platforms?

\textbf{RQ2: }How does the image aesthetic quality (IAQ) influence people’s engagement intentions with AIGP and HP?

\textbf{RQ3: }How are public sentiments toward AIGP? What factors motivate people's negative attitudes toward AIGP?
\vspace{12pt}

To answer these questions, we used human paintings as the baseline and compared the differences in user engagement levels and comment sentiments between AIGP and human paintings. To conduct the comparisons, we compiled a dataset with 207 AIGP and 210 human-painting \textit{image-text videos} and over 80,000 comments on these videos from TikTok. Image-text video is a format similar to PowerPoint, containing only images and text, where users swipe left or right to switch between images in the video. To answer RQ1, We used a formula to evaluate the user engagement level for every painting video. To answer RQ2, we adopted improved-aesthetic-predictor (LAION-Aesthetics V2\footnote{https://laion.ai/blog/laion-aesthetics/})
as the image aesthetic quality assessment (IAQA) model to assess the IAQ of each painting for further regression. To answer RQ3, we leveraged Natural Language Processing (NLP) techniques for sentiment analysis and topic modeling. The findings revealed that people still express stronger interaction intentions when browsing human paintings and that the IAQ does not significantly affect engagement intentions. We also found that people have more negative attitudes toward AIGP than human painting. The results of topic modeling listed seven main reasons for the negative sentiment. 


This paper makes contributions to the understanding of public perceptions towards AIGP on social media. Specifically, we contribute by:
\begin{itemize}
    \item Introducing image aesthetic quality (IAQ) as a critical moderating variable in the analysis of user engagement with AIGP and HP, thereby addressing a significant gap in the existing literature and providing novel insights into the dissemination patterns of AI-generated content on social media platforms.
    \item Conducting a comprehensive and large-scale analysis of public perceptions towards AIGP in the context of social media, based on user engagement metrics, comment sentiment scores, and topic modeling, using human paintings as a baseline for comparison.
    \item Identifying and categorizing seven key reasons behind negative public attitudes towards AIGP, including ‘Hyperrealistic Quality’, ‘Ambivalent Reactions’, ‘Failure to Meet Expectations’, and ‘Perceived Theft of Art’. These findings offer valuable guidance for the future optimization of AI generation tools to enhance user satisfaction and interaction experiences within human-computer interaction (HCI).
\end{itemize}


\section{Background and related work}
In this session, we briefly discuss the current background of AIGP technology development, what people think about the appearance of AIGP, and what the image aesthetic quality mentioned in this article is. We also provided loads of previous studies to support our research's objective and necessity.

\subsection{Social Attitudes toward Generative AI (GenAI)}

With the rapid development of AI, it has become increasingly important to study societal attitudes toward GenAI. Many researchers have also studied the attitudes of various groups of people towards generative AI technology in recent years. Grassini S et al.~\cite{grassini2023development} developed and validated an Artificial Intelligence Attitude Scale (AIAS) designed to assess public perceptions of AI technology. Vasiljeva T et al.~\cite{vasiljeva2021artificial}found that attitudes toward AI varied widely across industries. A significant difference in attitudes toward AI was found between employees of organizations implementing AI solutions and those without the intention of implementing AI solutions.
Pinto et al.'s~\cite{pinto2019medical} survey of 263 medical undergraduates found that undergraduate medical students were not concerned about AI replacing human radiologists, and they understood the potential applications and impact of AI on radiology and medicine. YILDIZ T et al.~\cite{yildiz2023measurement}found that language learners were delighted and favored using AI in the learning process. All in all, people seem to have different attitudes towards AI technology based on different usage scenarios and different populations.

AIGP, as the hot AI field of the moment, is developing at an astonishing rate, and they are already comparable to or surpass those created by humans. The technology not only brings us irreplaceable productivity and convenience~\cite{xu2019toward} but also the potential threat of indistinguishable deepfake problems~\cite{bregler2023video} and Substitution of work ~\cite{bendel2023image}. Therefore, to determine the direction of future AI technological development and people's needs, it is crucial and necessary to research the acceptance of artworks created by AI~\cite{colton2018issues}. 
Many scholars have conducted social surveys on people's attitudes towards AI-related work. Simon Colton~\cite{colton2008creativity} was concerned about the perceived rejection of creative computers, but the evidence was unclear. Moffat and Kelly~\cite{moffat2006investigation} proved that significant bias for computer-generated music existed. Martin Ragot et al.~\cite{ragot2020ai} invited 565 participants to rate paintings created by generative AI tools and humans, respectively, and the results presented that the scores of human paintings were significantly higher than AI paintings. However, the Painting quality question in this study has not been solved, and the generative tools were incomplete. So, few studies constructed a complete process to analyze the attitude toward AIGP, primarily focusing on the artwork circulating on social media.
In this research, we studied the social perception toward AIGP videos posted on social media and analyzed viewers’ sentiments from the interaction behaviors.

\subsection{Perception Differences between Human and AI Artwork}
  AI-driven images have demonstrated superior performance in producing a wide array of high-quality visuals, with some exhibiting richer color matching and more vivid detailing than human-created paintings. Given the similarity in visual effects, differences in people's perceptions and attitudes toward these two types of artwork have become a significant topic of discussion. Existing literature has analyzed this issue from various perspectives. For example, eye-tracking studies have revealed an implicit negative bias toward AI-generated art compared to human-created art, suggesting that artistic creativity is still predominantly perceived as a human attribute~\cite{zhou2023eyes}. Across two field experiments, researchers found that participants preferred artworks labeled as “human-created” over those labeled as “AI-created,” even when all artworks were, in fact, generated by AI~\cite{bellaiche2023humans}. Another experiment similarly revealed that art labeled as human-made was rated higher on aesthetic dimensions, particularly when directly compared to AI-made art, thereby enhancing the perceived value of human effort~\cite{horton2023bias}. To explore the underlying motivations behind these attitudes, one study showed that identical artworks were rated lower in terms of creativity and awe when labeled as AI-made rather than human-made. This labeling effect is one of the primary reasons for the reduced purchase intention toward AI-generated artwork~\cite{millet2023defending}.
  
  Despite the valuable insights provided by existing studies, several gaps remain. First, most research has focused on face-to-face surveys and controlled experiments, with limited attention given to user perceptions and attitudes toward AI-generated art and human-created art within the specific context of social media platforms. This is important, as certain features of social media—such as user-driven dissemination~\cite{kaplan2010users, wang2024pm} and the difficulty of verifying content accuracy~\cite{pennycook2019lazy}—may significantly influence users’ decision-making processes. Second, in many studies, the crucial attribute of aesthetic quality has not been adequately controlled or considered.

\subsection{Image Aesthetic Quality Assessment (IAQA)}

With the surge in visual artwork created by computers, we have witnessed AI-generated Paintings of varying quality~\cite{ramesh2021zero}. Meanwhile, it is increasingly challenging to filter and distinguish the chaotic and topic-unrelated images for some social media platforms. Therefore, how to efficiently and automatically evaluate the aesthetic quality of images spread on social media platforms has become an increasingly crucial problem. 

Aesthetics, a significant branch of the visual arts, delves into exploring aesthetic categories such as beauty and its counterpart, ugliness~\cite{zhu2010aesthetic}. The aesthetic appeal of an image is evaluated based on universally accepted principles of photography. This appeal can be influenced by various elements, such as the strategic use of colors~\cite{freeman2007complete}, the careful manipulation of contrast~\cite{itten1975design}, and the thoughtful arrangement of the image composition. Chatterjee and Leder et al.~\cite{leder2004model} introduced a five-stage, multi-level information processing model for image aesthetics. This model encompassed the perception of low-level information such as color, contrast, and complexity, the implicit integration of personal experiences and memories, explicit classification, cognitive mastery, and evaluation, culminating in aesthetic judgment and the generation of aesthetic emotions. However, computationally modeling this sequence for visual art images presents a significant challenge. Subsequently, Microsoft Asia Research Institute and Tsinghua University collaborated on a method to automatically differentiate photographs captured by professional photographers from those taken by customers~\cite {tong2005classification}. This study is widely recognized as the earliest research in the field of IAQA. The researchers employed a set of 21-class, 846-dimensional, low-level global features to train a classification model for aesthetically categorizing the test images. Until 2014, with the emergence of AVA, a large-scale image aesthetic analysis dataset, image aesthetic analysis has been given an enormous boost based on deep learning technology~\cite{jin2019ilgnet}. Lu X et al.~\cite{lu2014rapid} employed a dual-channel deep neural network to process local and global image blocks separately. Kong S et al.~\cite{kong2016photo} developed the AADB aesthetic database, rated by human evaluators, and utilized deep CNN for aesthetic ranking and categorization of images. Chang KY et al.~\cite{chan2019everybody} created an aesthetic dataset, PCCD, and introduced a CNN and LSTM-based deep neural network capable of generating aesthetic linguistic comments on images. Jin X et al.~\cite{jin2019ilgnet} designed a novel deep convolutional neural network, ILGNet, for distinguishing between images of high and low aesthetic quality. Some famous datasets, such as AVA (Aesthetic Visual Analysis)~\cite{murray2012ava}, AADB (Aesthetics and Attributes Database )~\cite{kong2016photo}, and CUHK-PQ (CUHK-Photo-Quality)~\cite{ke2006design} have collected billions of image samples. Therefore, using the aesthetic quality assessment models trained based on these datasets to evaluate the IAQ has been a good choice.

\section{Data and Methodology}
We integrated several quantitative methods to solve and answer the proposed research questions. We compared user engagement and comment sentiment between 207 image-text videos related to AIGP and 210 image-text videos showing human paintings. Additionally, to identify prominent reasons for negative comments, topic modeling was adopted to analyze the 42199 comments for AIGP and 39763 comments for human paintings, respectively.

\begin{figure*}[t]
\centering
\includegraphics[width=\textwidth]{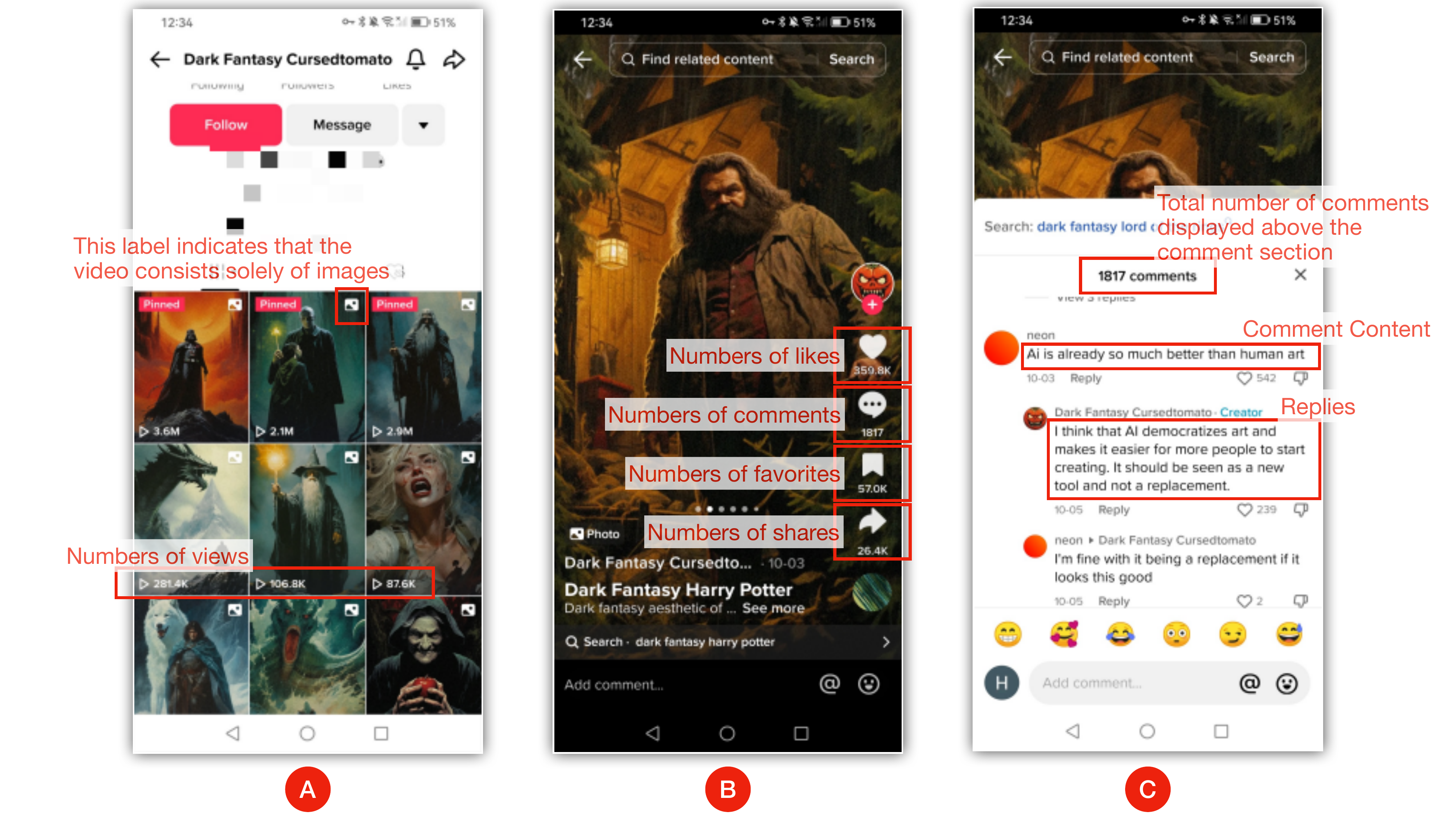} 
\caption{Overview of the user interface elements in the video platform. 
\textbf{A) }The creator's profile page on TikTok displays AIGI videos, showcasing thumbnails, view counts, and a label indicating that the video consists solely of images. 
\textbf{B) }The “For You” page highlights a selected video, presenting its associated metrics, including the number of likes, comments, favorites, and shares. 
\textbf{C) }The comments section contains AIGI-related discussions, showing the total number of comments, individual comment content, likes, and replies.}
\label{user interface}
\end{figure*}

\begin{figure}[t]
\centering
\includegraphics[width=1\columnwidth]{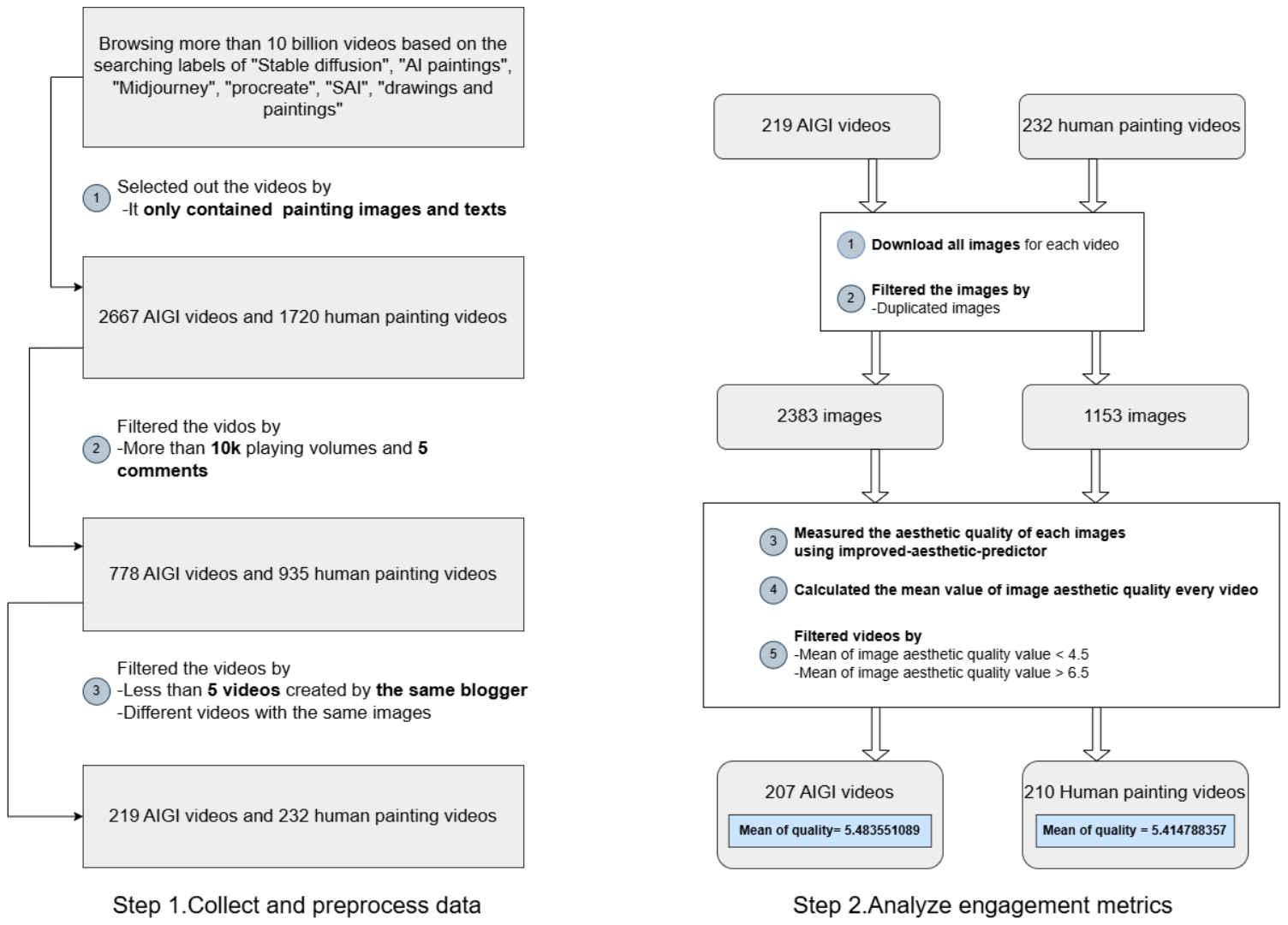} 
\caption{The data filtration process for AIGI and human painting videos. The \textbf{left side (Step 1)} illustrates the process of video selection and filtering based on content type, playing volume, and user engagement. 
The goal is to gather AIGI-related videos and ensure the dataset is clean and relevant for analysis. Videos are filtered using hashtags like ``stable diffusion'' or ``AI-generated images,'' and metadata such as view counts and upload dates are recorded. The \textbf{right side (Step 2)} demonstrates the aesthetic quality evaluation of extracted images and video categorization into four quality bins. This step focuses on assessing user interactions, such as likes, comments, and shares, to understand the popularity and reach of AIGI videos.}
\label{data filtration}
\end{figure}

\subsection{Hypotheses}
Previous research has indicated that people generally hold negative attitudes toward AI-generated content, particularly in the realm of artwork~\cite{hong2019artificial, ragot2020ai}. Therefore, our Hypotheses 1 and 3 are also grounded in this perspective, although we have narrowed the scope specifically to AI-generated paintings~\cite{wu2020investigating}. Meanwhile, visual aesthetic quality has been shown to be a significant factor in promoting user behaviors, such as downloading and rating items on the Internet~\cite{bhandari2019understanding}.

In addition to the direct effect of IAQ on user engagement, we consider that IAQ may also play a moderating role in how users evaluate content from different sources, as suggested by existing literature. One study indicated that perceived aesthetics moderated trust and satisfaction in e-commerce environments~\cite{cyr2010colour}. In the context of browser-based decision making, another study revealed that high aesthetic quality amplifies positive responses~\cite{reinecke2013predicting}. Applying this mechanism to the context of our study, a higher level of IAQ may help offset users’ inherent bias against AI-generated artwork, leading viewers to develop greater acceptance of AI art and, consequently, encouraging interactive behaviors. To facilitate understanding of the moderating model in this study, we use “human origin level" to represent the creator type, with 0 defined as AIGP and 1 as HP. Therefore, we propose Hypothesis 3 that IAQ may negatively moderate the relationship between human origin level and user engagement. To answer these research questions, we proposed the following three hypotheses:

\vspace{12pt}
\textbf{H1: }People are more willing to interact with human paintings than with AIGP on social media platforms.

\textbf{H2: }IAQ plays a negative moderating role in the relationship between human origin level and user engagement.

\textbf{H3: }People express more negative emotions in comments on AIGP than human paintings on social media.

\subsection{Research Contexts}
This study is based on user-generated content (UGC) from social media platforms. The choice of social media as the research context is primarily based on two main reasons. Firstly, social media provides real-time communication channels, enabling researchers to capture immediate and authentic interactions and perceptions from users. According to the Authentic Self Theory, individuals tend to express themselves more freely on social media, as they are less constrained by traditional societal expectations and norms. This relatively anonymous and disinhibited environment encourages users to present a more authentic self and to express their genuine attitudes~\cite{vannini2008authenticity}. Secondly, social media brings together users from diverse demographic and geographic backgrounds, allowing researchers to effectively analyze broad public opinions and perceptions~\cite{perrin2019share}. This helps mitigate potential biases, such as regional differences. Moreover, social media offers access to large-scale datasets, allowing for more comprehensive and reliable analysis. Therefore, social media platforms are highly suitable as the research context for this study.

In addition, among many social media platforms, TikTok has become the fastest-growing social network in the post-pandemic era, allowing users to publish and share short videos ranging from 15 to 60 seconds. It recorded 1.506 billion downloads in 2021, surpassing Instagram, which had 1.048 billion downloads~\cite{barta2023influencer}. In 2022, the average daily time users worldwide spent on the TikTok mobile app was nearly equivalent to the combined time spent on Facebook and Instagram~\cite{zhang2023social}. 

There are three main reasons for choosing TikTok as the survey platform for this study. First, as of December 15, 2023, videos under tags related to AI painting on TikTok, such as ``aiart'', ``midjourney'', and ``stable diffusion'', had accumulated 10.7B, 7.5B, and 2.3B views, respectively. This indicates that a large amount of AIGP-related content has been widely disseminated on TikTok, and that users show a high level of motivation and willingness to engage in discussions on AI-related topics. Second, TikTok’s recommendation algorithm analyzes users’ viewing preferences and pushes content directly to them on the ``For You'' page~\cite{zhao2020analysis}. In this process, users are unaware of what video will appear next, which reduces the influence of utilitarian behavior on video engagement metrics in the experimental data. According to the Authentic Self Theory, individuals are more likely to exhibit genuine reactions when confronted with random events, as they have limited time to prepare or control their responses~\cite{vannini2008authenticity}. Therefore, when a video with unpredictable content appears suddenly on TikTok, the user's reaction is more likely to be authentic. Last but not least, according to TikTok’s platform management guidelines, all videos involving AI-generated content must be labeled. Videos without proper labeling will not pass the platform's review. Fig.~\ref{user interface} shows an overview of TikTok’s user interface elements. Therefore, compared with other social media platforms such as Reddit, Instagram, and Twitter, TikTok is undoubtedly the most appropriate choice.

\subsection{Data Collection and Pre-processing}
In our preliminary survey, we discovered that paintings on TikTok are presented in two formats: short videos and image-text videos. In short videos, many creators share their motivation for creating, record the creative process, or provide step-by-step painting tutorials. In these videos, user engagement is influenced by many factors unrelated to the artwork itself, such as the creator's voice and behavior, the video's length, etc.~\cite{yang2022science}, which means that it is difficult to capture people's actual attitudes toward the two types of paintings from these videos. Therefore, we selected image-text videos as our experimental objectives, which are similar to PowerPoint format and contain only images and text. Users swipe left or right to switch between images in the video. In this kind of video, creators act more as sharers than narrators, only sharing their creations or AI-generated paintings. Thus, user engagement depends mainly on their understanding and feelings towards the artwork, unaffected by the creator's value output. Therefore, choosing image-text videos for comparative analysis perfectly aligns with our research theme, making our experimental results more scientific and reliable.

We created a new account on the TikTok video-sharing social media platform specifically for this study. By searching for tags related to AIGP and human paintings on TikTok, such as ``Stable Diffusion'', ``Midjourney'', ``AI paintings'', ``Procreate'', and ``SAI'', and by scraping the web, we selected 2,667 AI-labeled image-text videos and 1,729 human painting image-text videos as of November 15, 2023. The selection of these search terms was guided by their direct relevance to the intersection of AI technology and digital art creation. They encompass both the AI tools used to generate art and the platforms where such art is further refined or utilized, ensuring a comprehensive exploration of public interactions and perceptions related to AI-generated paintings.

Considering that videos with low view counts might not be representative and that some creators may restrict comment permissions, we excluded videos with fewer than 10,000 views, videos irrelevant to painting, and those with fewer than five comments. There are two main reasons for this exclusion. First, videos with fewer than five comments do not provide sufficient data to support subsequent sentiment and semantic analyses. Second, video view counts on TikTok are largely influenced by the platform’s recommendation algorithm~\cite{klug2021trick}. The biases inherent in this algorithm may lead to a mismatch between the actual quality of a video and its level of exposure. For example, some high-quality videos may not receive adequate visibility, resulting in incomplete or unreliable user-related data. This, in turn, could interfere with the accuracy of our data analysis. Therefore, to minimize the potential impact of algorithmic mechanisms on the experimental results, such videos were excluded from the dataset. This filtering process refined our dataset to 778 AIGP videos and 935 human painting videos. To avoid user favoritism or bias toward specific creators and to prevent the inclusion of repetitive paintings, we adopted a same-author avoidance principle. Specifically, we limited the sample to no more than five works per author and excluded videos containing repetitive paintings within the same video.

Finally, after applying all the filtering criteria, we obtained a dataset comprising 207 AIGP image-text videos and 210 human painting image-text videos. The holistic data collection and filtration process is illustrated in Fig.~\ref{data filtration}.

\begin{figure}[H]
\centering
\includegraphics[width=0.8\columnwidth]{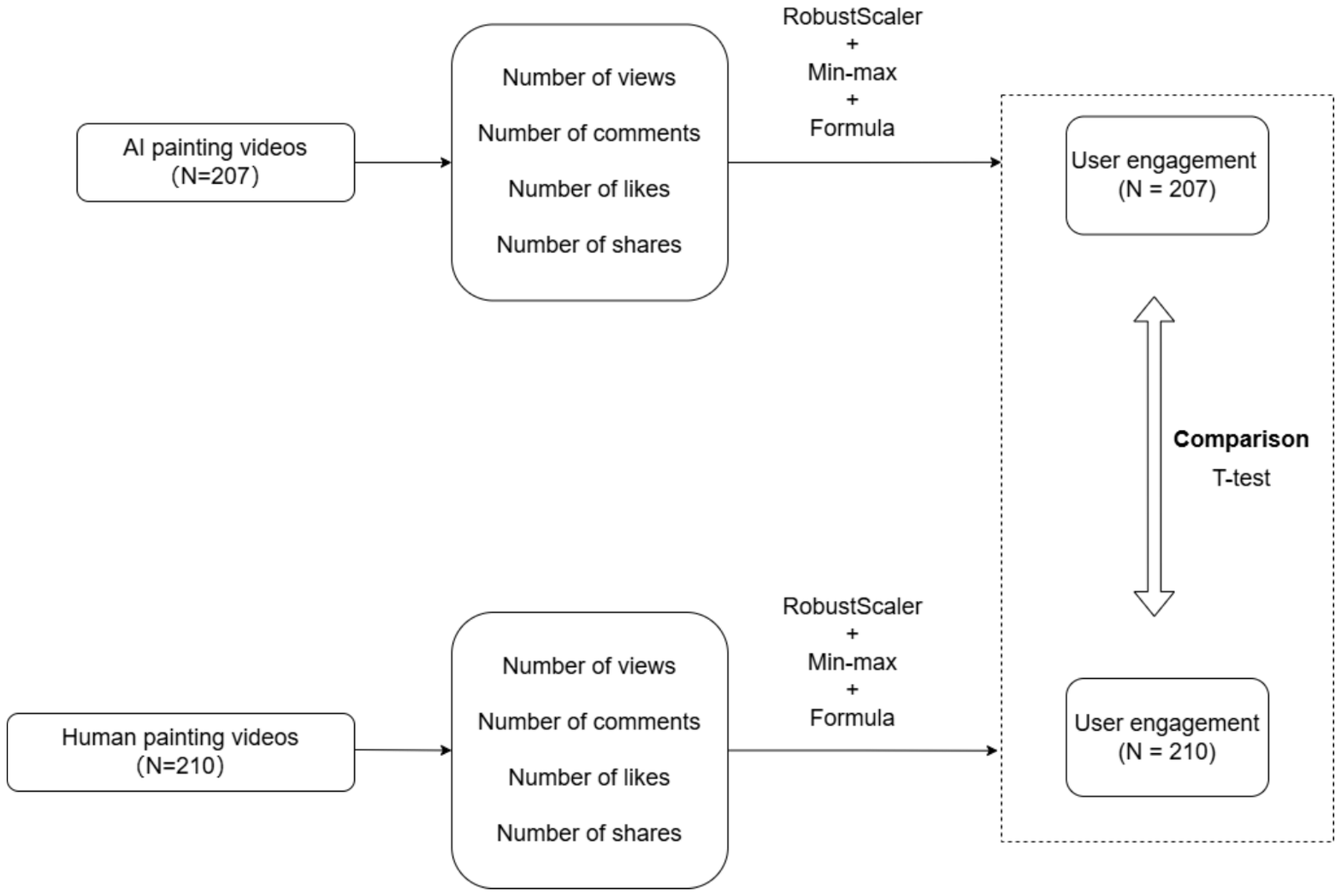} 
\caption{Flowchart for calculating and comparing user engagement. 
This flowchart illustrates the process of calculating user engagement for AI-generated painting videos (N = 207) and human painting videos (N = 210). Engagement metrics, including number of views, number of comments, number of likes and number of shares, were normalized using RobustScaler and Min-Max scaling. The resulting user engagement scores were compared using the Independent T-test.}
\label{Flowchart_calculating}
\end{figure}

\subsection{User Engagement Analysis (RQ1)}

In this work, we define user engagement as the interactivity between users and videos~\cite{o2008user}, reflecting the audience's genuine interest and interaction with the videos. We adopted a formula for calculating TikTok user engagement on the HypeAuditor platform, an AI-driven analytical and discovery tool for Instagram, YouTube, TikTok, Twitter, and Twitch. It helps to find relevant influencers, understand their audience, increase the ROI (Return On Investment) of advertisers, and maintain the authenticity of influencer marketing. Many previous studies on social media data analysis have used this tool\cite{trevisan2019towards,de2022avatar}. The formula for TikTok user engagement on Hype is $[(number of likes + number of comments + number of shares) / number of views] X 100\% $ \footnote{https://hypeauditor.com/blog/what-is-tiktok-engagement-rate-why-brands-should-take-note/}. Previous studies on TikTok engagement have also adopted these three metrics\cite{wang2024critical}. The flowchart for calculating and comparing user engagement is shown in Fig. \ref{Flowchart_calculating}

\subsection{Image Aesthetic Quality Assessment (IAQA) (RQ2)}

Video quality is an important control variable in this paper, and its impact on user engagement has been confirmed by many previous studies\cite{diallo2014impacts}. So, in the process of sample preprocessing, we also considered the video quality. Since image-text videos only contain Paintings and text, all videos have the same page layout and presentation method, so here, video quality is equivalent to the aesthetic quality of the Paintings in the videos. Firstly, we downloaded all the Paintings from the two types of videos, removed duplicate Paintings within the same video, and obtained 2383 AI paintings and 1153 human paintings. In our research, we used the improved-aesthetic-predictor (LAION-Aesthetics V2)\footnote{https://laion.ai/blog/laion-aesthetics/} as the IAQA model. The model's training process is divided into two steps: embedding the input images using Clip ViT/14~\cite{radford2021learning} and then using MLP to train the image data from three well-known datasets, namely, 176,000 image-text pairs from SAC~\cite{pressman2023simulacra} dataset, 15,000 image-text pairs LAION-Logos dataset with aesthetic scores from 1 to 10, and 250,000 photos from the AVA dataset with aesthetic scores from 1 to 10. The data after aesthetic quality scoring of 2.73 billion images in the LAION 5B dataset~\cite{schuhmann2022laion} by this Predictor trained Stable Diffusion V1\footnote{https://github.com/CompVis/stable-diffusion/tree/ce05de28194041e030ccfc70c635fe3707cdfc30\#stable-diffusion-v1}, one of the most professional AIGC models currently. Therefore, this model's comprehensive reliability and accuracy in assessing image aesthetic quality are likely the best now available.

\begin{figure}[t]
\centering
\includegraphics[width=1\columnwidth]{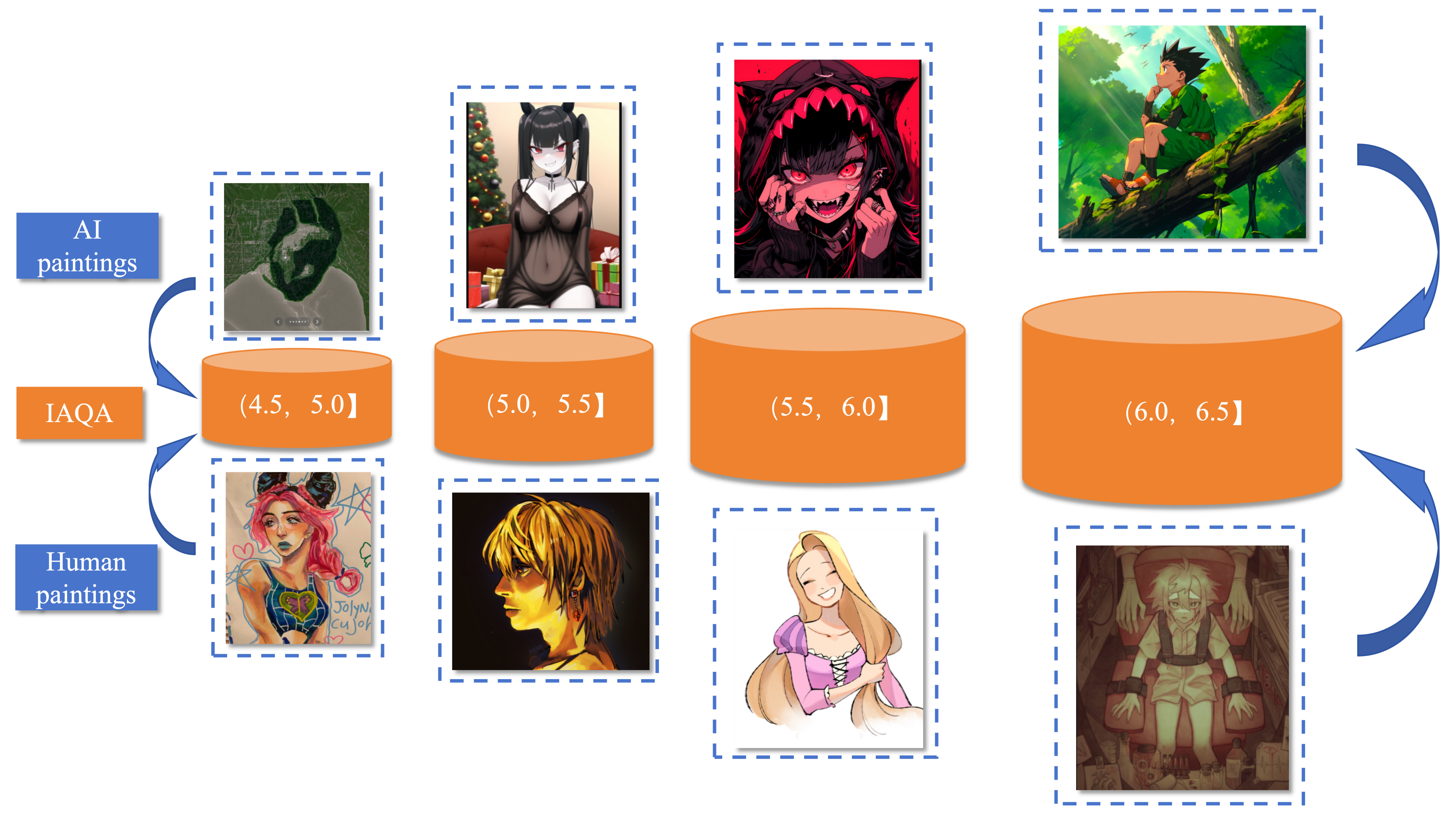} 
\caption{Image Aesthetic Quality Schematic. This schematic illustrates images from both categories that were evaluated using the Image Aesthetic Quality Assessment (IAQA) tool, which assigned each image a quality score.}
\label{fig3}
\end{figure}

\begin{table*}[ht]
\centering
\resizebox{0.9\textwidth}{!}{ 
\begin{tabular}{ccp{10cm}}
\noalign{\hrule height 0.4mm}
\textbf{} & \textbf{Variables} & \textbf{Description} \\
\noalign{\hrule height 0.4mm}
Dependent variable & User Engagement (Ue) & User Engagement refers to the level of user interaction with the content. In our study, we measure this variable using the number of comments, likes, and shares. \\
\hline
Independent variable & Human Origin Level (Hum) & Human origin level is a binary variable indicating the origin of a given painting. It is coded as 0 for AIGP and 1 for HP \\
\hline
Moderator & Aesthetic Quality (Aes) & Image Aesthetic Quality represents the aesthetic quality of the painting. We measure this variable using aesthetic quality scores. \\
\noalign{\hrule height 0.4mm}
\end{tabular}}
\caption{Descriptions of variables in the regression model}
\label{table:description}
\end{table*}

After collecting and measuring the IAQ of approximately 3,500 images of AIGP and HP published on the TikTok platform, we found that the IAQ scores of both types of paintings predominantly ranged between 4.5 and 6.5. To ensure comparability and avoid the influence of extreme values on the experimental results, we excluded videos with an average image aesthetic quality score lower than 4.5 or higher than 6.5. This decision was made for two main reasons: (1) extremely low or high scores often corresponded to outlier cases—such as distorted, poorly generated, or overly edited images—which may introduce noise and distort the analysis; and (2) the majority of both AIGP and HP samples were concentrated within this score range, making it a representative and comparable interval for analysis. After applying this filtering criterion, we obtained 207 AIGP videos and 210 HP videos for further analysis. We illustrate some Paintings from different aesthetic quality intervals in Fig.\ref{fig3}. By counting the total number of views of the two types of videos, we find that their total number of views has reached hundreds of millions, which indicates that a large number of users are viewing similar videos on TikTok. Therefore, our sample data is universal and reliable. 

To test H2, we built an OLS regression model shown below and used Stata to examine whether IAQ moderates the relationship between user engagement and painting sources. Table \ref{table:description} shows the descriptions of each variable in the regression model. The dependent variable is User Engagement (\textit{Ue}), which captures the level of user interaction with each video. This variable is operationalized using observable engagement metrics such as the number of likes, comments, and shares. The independent variable is Human Origin Level (\textit{Hum}), which distinguishes between two types of visual content: AIGP as 0 and HP as 1. This variable allows us to examine differences in engagement based on the origin of the artwork. In addition, Image Aesthetic (\textit{Aes}) is introduced as a moderator variable, reflecting the aesthetic quality of each image. This variable is measured using quantitative aesthetic quality scores and is used to test whether aesthetic appeal moderates the relationship between painting source and user engagement. These variables collectively form the basis for exploring how the source and aesthetic quality of artwork jointly influence user behavior on social media platforms.

Since all data cleaning and preprocessing procedures were already completed in the “Data Collection and Pre-processing" section, we do not repeat those steps here. However, to reduce potential multicollinearity and enhance the stability of the regression analysis, we applied mean-centering to the continuous moderator variable, \textit{Aes}. Centering was performed by subtracting the sample mean from each observed value of the variable, as shown below:

\begin{equation}
\text{Aes}_{\text{centered}} = \text{Aes}_i - \overline{\text{Aes}}
\end{equation}

\vspace{0.2cm}

Finally, we build the following OLS regression model and analyze it using state, as shown below:

\vspace{0.4cm}

\begin{equation}
Ue = \beta_0 + \beta_1 \cdot \text{Hum} + \beta_2 \cdot \text{Aes}_{\text{centered}} + \beta_3 \cdot \text{Hum} \cdot \text{Aes}_{\text{centered}} + \varepsilon_i
\end{equation}

\subsection{Sentiment Analysis and Topic Modeling (RQ3)}

\begin{figure}[ht]
\centering
\includegraphics[width=0.6\columnwidth]{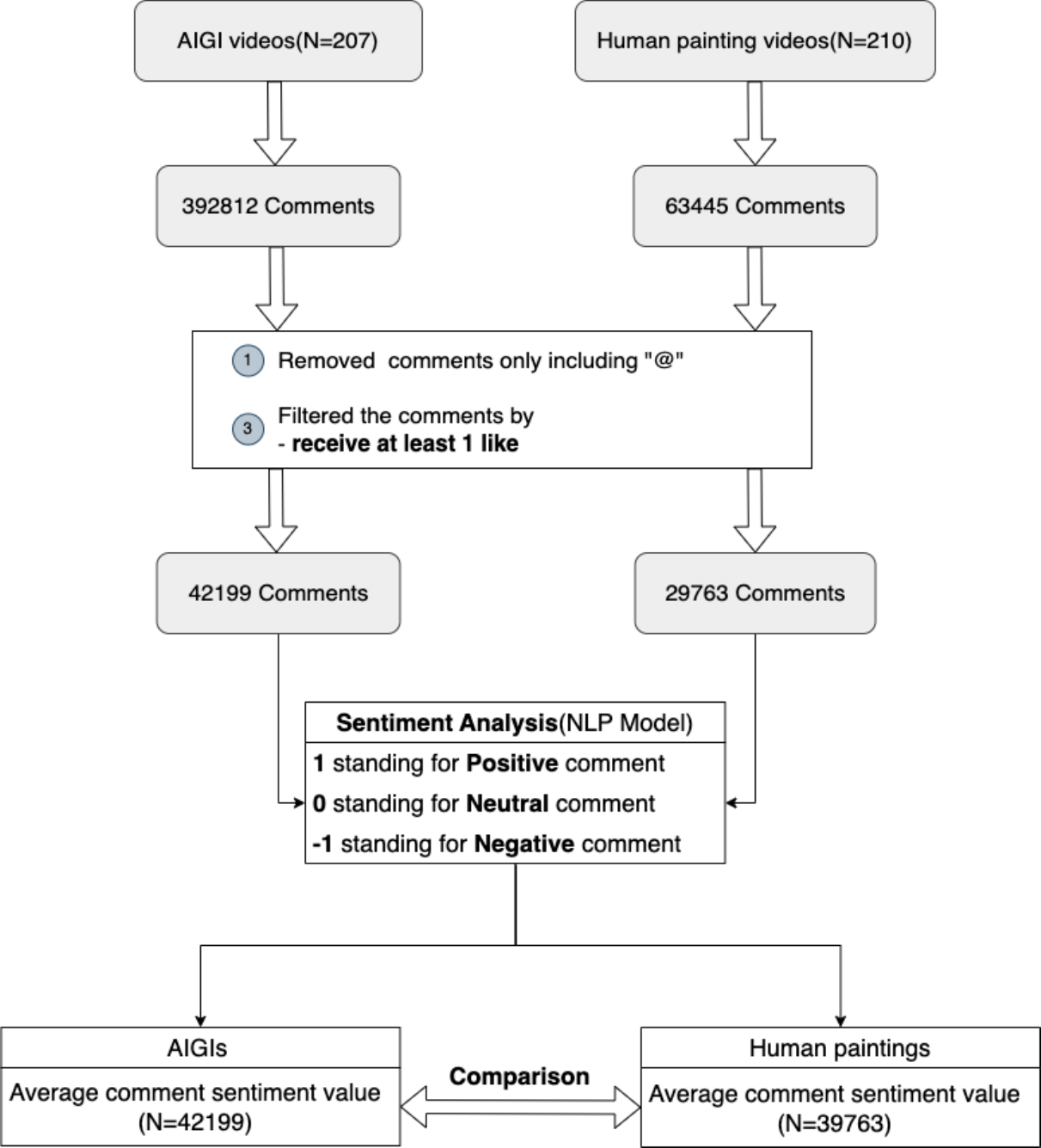} 
\caption{Sentiment Analysis Flowchart.
This flowchart demonstrates the process of performing sentiment analysis on comments from AIGP and HP. Initially, 392,812 comments from AIGI videos and 63,445 comments from human painting videos were collected. Comments containing only ``@'' were removed, and comments with at least one like were retained, resulting in 42,199 comments for AIGI videos and 29,763 comments for human painting videos. Sentiment analysis was conducted using an NLP model, where comments were classified as positive (1), neutral (0), or negative (-1). The average sentiment values for AIGI and human painting videos were calculated and compared.}
\label{Sentiment_analysis}
\end{figure}

In this study, sentiment analysis was conducted on comments from two types of videos to reflect the genuine inner feelings of users, thereby revealing people's real attitudes towards AIGP on social media. Fig.\ref{Sentiment_analysis} shows the flowchart of comment sentiment analysis and comparison. We extracted all comments from the sample videos, resulting in 392,812 comments for AIGP videos and 63,445 for Human painting videos. First, we used the open-source model 'Hello-SimpleAI/chatgpt-detector-roberta'~\cite{guo2023close} on Hugging Face to detect if any comments were generated by bots, and then we removed those comments. Comments that were only ``@'' or had received no likes were removed too. After filtering, 42,199 comments from AIGP videos and 29,763 comments from Human painting videos were retained. These comments were then inputted into a sentiment analysis model named ``distilbert-base-multilingual-cased-sentiments-student'' \footnote{https://huggingface.co/lxyuan/distilbert-base-multilingual-cased-sentiments-student}. It's an open-source model based on distilbert~\cite{sanh2019distilbert} training designed for multilingual applications, capable of recognizing twelve languages and interpreting sentiments as positive, neutral, or negative. As of December 30, 2023, it has been downloaded over 7 million times on Hugging Face. Since the model recognizes only twelve languages, ChatGPT 4.0 was used to translate the comments outside of the twelve languages into English. The model's output categorized positive sentiment comments as 1, neutral comments as 0, and negative comments as -1. Since the final results did not conform to a normal distribution, the Wilcoxon test was chosen to compare the sentiment scores of the two types of videos. 

\begin{table*}[t] 
\centering

\scalebox{0.88}{
\begin{tabular}{cccccccc}
\hline
\textbf{Measures} & \textbf{Normality} & \textbf{Test types} & \textbf{Items} & \textbf{Mean} & \textbf{SD} & \textbf{t}  & \textbf{p}\\ \hline
\multirow{2}{*}{User engagement} & \multirow{2}{*}{N} & \multirow{2}{*}{Independent T-test} & AIGP videos (N = 207) & 0.104 & 0.058 & \multirow{2}{*}{-15.064} & 0.00*** \\
 & & & HP videos (N = 210) & 0.198 & 0.068 & & \\ \hline
\end{tabular}}
\caption{The user engagement calculated by traditional formula. The normality (N) of the data was assessed using the Shapiro-Wilk test. We used an independent T-test if the difference is normally distributed and the Wilcoxon test otherwise.}
\label{table: user engagement}
\end{table*} 

\begin{table*}[t] 
\centering
\scalebox{0.97}{
\begin{tabular}{cccccccc}
\hline
\textbf{Measures} & \textbf{Normality} & \textbf{Test types} & \textbf{Items} & \textbf{Mean} & \textbf{SD} & \textbf{W}  & \textbf{p} \\ \hline
\multirow{2}{*}{Sentiment scores} & \multirow{2}{*}{N} & \multirow{2}{*}{Wilcoxon test} & AIGP videos (N = 207) & 0.3 & 0.93 & \multirow{2}{*}{-36.725} & \multirow{2}{*}{0.00***} \\
 & & & HP videos (N = 210) & 0.53 & 0.83 & & \\ \hline
\end{tabular}}
\caption{A comparison of the comment sentiment scores. The normality (N) of the data was assessed using the Shapiro-Wilk test. We used an independent T-test if the difference is normally distributed and the Wilcoxon test otherwise.}
\label{table: sentiment}
\end{table*} 

\begin{figure}[ht]
  \centering
  \includegraphics[width=\linewidth]{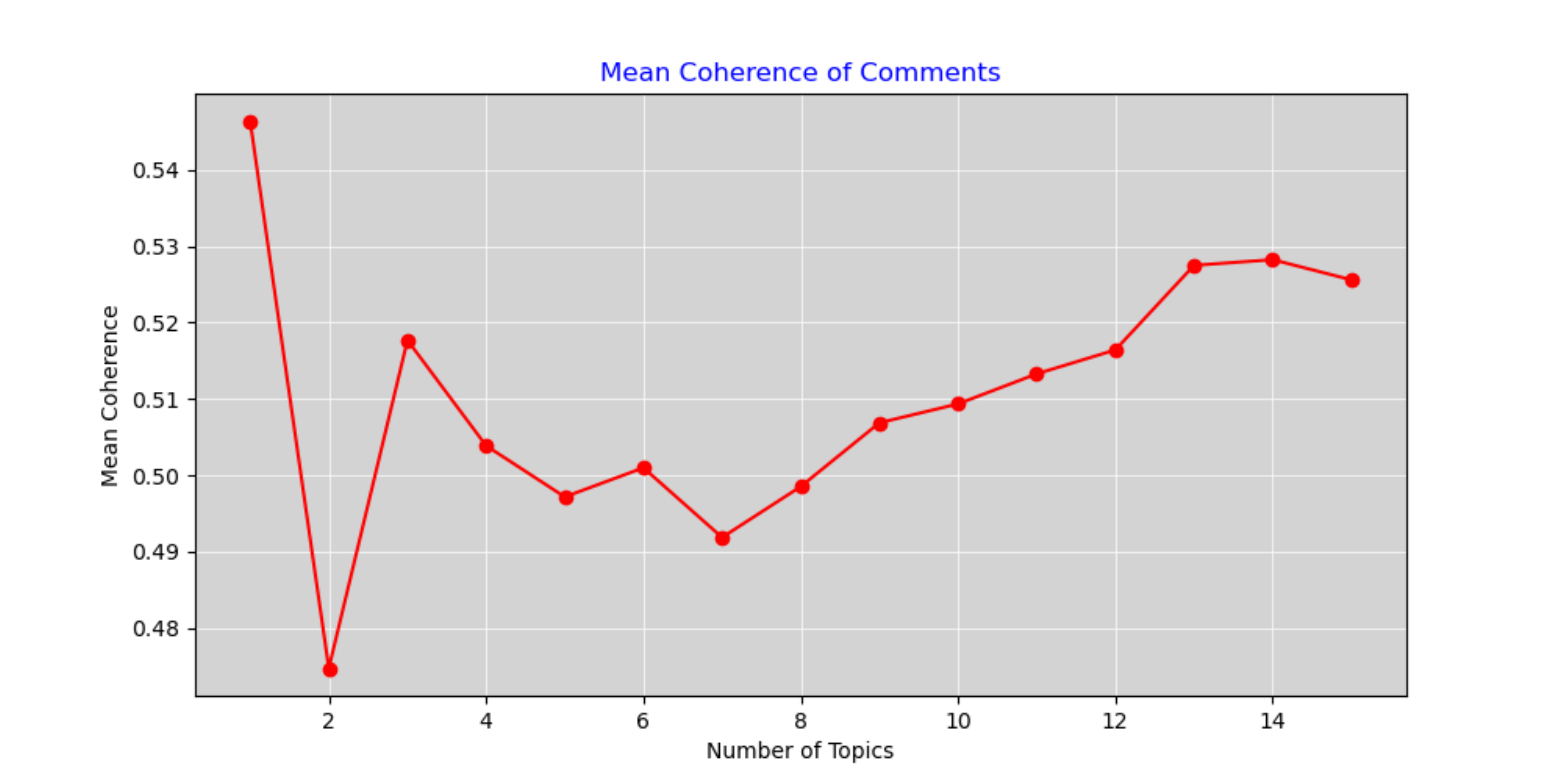}
  \caption{Mean Coherence of Comments.
This figure shows the mean coherence scores of comment topics as a function of the number of topics. }
  \label{fig:topic modeling}
\end{figure}

For topic modeling, we initially filtered the comments on AIGP videos to extract all negative comments, resulting in 19,071 comments. These comments were subjected to topic modeling in this study. After comparing several mainstream topic modeling methods, such as Latent Dirichlet Allocation (LDA), Latent Semantic Analysis (LSA), and Non-Negative Matrix Factorization (NMF), we opted for the NMF model because our data contained a number of short texts but not long documents~\cite {gamage2022deepfakes}. We trained the NMF model for different numbers of topics from \textit{k} = 1 to 15, and we calculated the mean topic coherence across all topics to determine the optimal number of topics automatically (Fig. \ref{fig:topic modeling}). Semantic coherence involves examining the co-occurrence of words within a document to ensure that the topics generated by the model possess semantic cohesion ~\cite{wang2023making}, a mainstream method for assessing topic quality. When \textit{k} = 1, semantic coherence was the highest, but a single topic could not accurately summarize our collected comment data. Hence, we chose 14 topics for the final training phase, as this number demonstrated a relatively high level of semantic coherence.

\section{Results}

To compare user engagement levels between AIGP and HP on TikTok, we conducted an independent samples t-test. Prior to the analysis, the normality of the user engagement data was assessed using the Shapiro-Wilk test. As the data met the assumption of normality, an independent t-test was deemed appropriate. The results reveal a statistically significant difference in user engagement between AIGP videos ($M = 0.104, Std = 0.058$) and HP videos ($M = 0.198, Std = 0.068$), $t(415) = –15.064, p < 0.001$ in Table \ref{table: user engagement}. The mean engagement for human-created paintings is substantially higher than that for AI-generated content. This finding suggests that users interact significantly more with human-created paintings compared to AIGP on social media platforms. These results indicate that H1 is accepted.

To examine whether IAQ positively moderates the relationship between Human Origin Level (Hum) and User Engagement (Ue), we conducted a regression analysis using Stata. The results are presented in Table \ref{table:regression}. The coefficient for \textit{Hum} is positive and statistically significant ($\beta = 0.094, p < 0.001$), indicating that HP is associated with higher levels of user engagement compared to AIGP, holding other variables constant. The coefficient for \textit{Aes} is positive but not statistically significant ($\beta = 0.008, p = 0.387$), suggesting that aesthetic quality alone does not have a significant direct effect on user engagement in this model. Crucially, the interaction term \textit{Hum × Aes} has a negative and non-significant coefficient ($\beta = –0.007, p = 0.620$), indicating that IAQ does not significantly moderate the relationship between \textit{Hum} and \textit{Ue}. In other words, the effect of \textit{Hum} on user engagement remains consistent regardless of variations in aesthetic quality. The overall model explains approximately 35.5\% of the variance in user engagement ($ R^2 = 0.355; Adjusted R^2 = 0.350$), suggesting a moderate level of explanatory power. In summary, while the type of painting source significantly influences user engagement, there is no evidence from this model that IAQ moderates this relationship. Hence, the experimental results indicate that hypothesis H2 is rejected.

To compare user comments on AI-generated paintings (AIGP) versus human-made paintings (HP) on social media, we conducted semantic analysis, including sentiment score comparisons and topic modeling based on the collected comments. Table \ref{table: sentiment} presents the results of the sentiment analysis. As the normality assumption was violated (as assessed by the Shapiro-Wilk test), a Wilcoxon rank-sum test—the non-parametric equivalent of the independent t-test—was employed. The results indicate a statistically significant difference in sentiment scores between the two groups. Specifically, comments on AIGP videos ($N = 207$) had a mean sentiment score of 0.30 ($SD = 0.93$), while comments on HP videos ($N = 210$) had a higher mean sentiment score of 0.53 ($SD = 0.83$). The Wilcoxon test yielded a test statistic of $W = -36.725$ with a p-value less than 0.001, indicating a highly significant difference. These results suggest that comments on HP content tend to be significantly more positive than those on AIGP content. In other words, users express more favorable sentiment toward content featuring human-made paintings compared to AI-generated ones, highlighting potential perceptual or emotional biases in user responses. Based on the sentiment analysis, hypothesis H3 is rejected.

Secondly, the mean coherence scores and the results of topic modeling are presented in Figure~\ref{fig:topic modeling} and Table~\ref{table:topic_modeling}. Figure~\ref{fig:topic modeling} shows a non-linear trend in coherence scores across different numbers of topics. Although the highest coherence score occurs at two topics, this result is often considered unreliable due to its limited interpretability and the overly broad grouping of themes. After a sharp decline at three topics, the coherence scores gradually increase as the number of topics rises, reaching a local peak at 14 topics. Based on these findings, we selected the model with 14 topics for further qualitative analysis, as it offers a balance between coherence and interpretability. From this model, we identified the top seven topics with the highest proportions, which are presented in Table~\ref{table:topic_modeling}.

\begin{table}[ht]
\centering

\begin{tabular}{>{\centering\arraybackslash}p{3.5cm}|>{\centering\arraybackslash}p{3.5cm}}
\noalign{\hrule height 0.4mm}
\textbf{Variables} & \textbf{Model} \\
\noalign{\hrule height 0.4mm}
\multirow{2}{*}{Hum} & 0.094\textsuperscript{***} \\
& (0.000) \\
\hline
\multirow{2}{*}{Aes} & 0.008 \\
& (0.387) \\
\hline
\multirow{2}{*}{Hum*Aes} & -0.007 \\
& (0.620) \\
\hline
Constant & 0.104 \\
\hline
R-squared & 0.355 \\
\hline
Adj R-squared & 0.350 \\
\noalign{\hrule height 0.4mm}
\end{tabular}
\caption{Table 4: Regression results. The results shows that while Human Origin Level affects user engagement, IAQ does not moderate this relationship. Note: *p\textless0.05, **p\textless0.01, ***p\textless0.001}
\label{table:regression}
\end{table}

\begin{table*}[ht] 
\centering
\resizebox{0.8\textwidth}{!}{
\begin{tabular}{ccccc}
\hline
\textbf{Items} & 
\textbf{\begin{tabular}[c]{@{}c@{}}Like-to-view \\ ratio\end{tabular}} & 
\textbf{\begin{tabular}[c]{@{}c@{}}Comment-to-view \\ ratio\end{tabular}} & 
\textbf{\begin{tabular}[c]{@{}c@{}}Favorite-to-view \\ ratio\end{tabular}} & 
\textbf{\begin{tabular}[c]{@{}c@{}}Share-to-view \\ ratio\end{tabular}} \\

\hline
AIG videos (N=207) & 0.10107 & 0.00101 & 0.02057 & 0.00224 \\
HP videos (N=210) & 0.19388 & 0.0029 & 0.02779 & 0.00081 \\
\hline
\end{tabular}}
\caption{Like ratio, comment ratio, favorite ratio, and share ratio}
\label{table: ratio}
\end{table*}

\section{Findings}
Based on the results of the data analysis, this section provides detailed answers to each of the research questions proposed in this study.

\subsection{RQ1: How are people's engagement levels with AIGP on social media platforms?}
As seen in Table \ref{table: user engagement}, when the quality of the video content is the same, people are more willing to interact with human painting videos than with AIGP videos. Because our experimental data collection and pre-processing process has eliminated the interference of most other variables, in such a situation, the response made by the user to the video can almost be equated with the user's real attitude towards the paintings. So, the users' behaviors and decisions are solely based on the paintings, such as liking, commenting, favoriting, and sharing. We can see that although there is a lot of AIGP-related content on social media and there are many people who have viewed AIGP works, users still prefer to browse and interact with human paintings on social media platforms.

\subsubsection{Stronger Desire to Share AIGP}

We calculated the like, comment, favorite, and share ratios of each video based on views in Table \ref{table: ratio}. We found that the like, comment, and favorite ratios for human painting videos significantly exceed those for AIGP videos, particularly in the comment ratios of 0.001 (AIGP) and 0.003 (HP). These data corroborate our previous results on user engagement comparison, and also show that using machine learning methods to calculate user engagement is reliable. Interestingly, the share ratio of AIGP videos is much higher than that of human painting videos. Based on our speculation, the possible reason is that the ability of AI painting breaks people's perception of AI technology, and the speed of its development is also shocking to people who want to share it and discuss it with their friends. However, as with comments, we are not clear about users' motivations for sharing, which can be positive or negative signals. Therefore, we analyzed the comment motivation specifically in a later section.

\subsection{RQ2: How does the image aesthetic quality (IAQ) moderate people’s engagement intentions with AIGP?}

We analyzed the potential moderating role of IAQ on the relationship of user engagement and painting sources, and the final results found no significant influence. This suggests that when browsing different artworks on social media, people may be more concerned with the intrinsic meaning and value of the paintings than their glossy appearance. Moreover, we noticed that the engagement with human painting videos increases slightly as aesthetic quality decreases. Observing the comments on human painting videos, there appears to be a lot of positive social support for the creators, such as when amateur painters upload less-than-perfect paintings and drawings on TikTok. Interestingly, many users offer encouragement rather than criticism. Some examples are as follows:

\begin{quote}
    ``\textit{The Drawing Looks Better Than The Original!}''
    
    ``\textit{your art just gets better and better I’m amazed!}''
    
    ``\textit{YOUR OLD ART IS SO PRETTY BUT YOUR CURRENT ART IS EVEN BETTER!! You've definitely improved I'm obsessed with your style.}''
\end{quote}
Such emotional support (e.g., encouragement, intimacy, compliment, and empathy) can give individuals a sense of respect and achievement, strengthening community ties and member engagement~\cite{wang2012stay}. At the same time, viewer behaviors may also be influenced by the herd effect~\cite{yen2023crowdidea}, following the content of the majority's comments and involuntarily giving a steady stream of support to the art creator. This emotional value then stimulates the creator's willingness to create. So this may be one of the reasons why the lower the aesthetic quality of the Painting, the higher the engagement of human painting videos. But we don't find such a large-scale phenomenon in AI painting videos.

\begin{table*}[t]
\centering
\resizebox{0.8\textwidth}{!}{%
\begin{tabular}{p{0.5cm} p{4cm} p{7cm} p{3cm} p{2cm}}
\toprule[1.2pt]
\textbf{\#No} & \textbf{Topics} & \textbf{Description} & \textbf{Keywords} & \textbf{Proportion} \\
\toprule[1.2pt]
1 & Hyperrealistic Quality & This topic reflects users' perceptions that AI-generated images exhibit a hyperrealistic quality, making them almost indistinguishable from traditional artwork created by human painters. While many users express admiration for the technical precision and realistic details, others feel discomfort or unease, describing the images as “too real” to the point of being unnatural or unsettling. & like, looks, artist, sick, cute, real, bad, feel, movies, girl, style & 12.82\% \\
\cmidrule{1-5}
2 & Ambivalent Reactions & This topic reflects users' shift from initial admiration for the artwork’s visual quality to criticism upon discovering it was AI-generated. The transition is driven by concerns about originality, authenticity, and the creative value of AI art, highlighting a tension between appreciation of its technical brilliance and skepticism about its legitimacy as art. & damn, hate, shame, fuck, cute, insane, great, disappoint, interesting, cool, pretty & 12.72\% \\
\cmidrule{1-5}
3 & Rejection of AI Art & This topic represents the outright refusal to recognize or accept AI-generated art as legitimate. Users often criticize such works for lacking emotional depth, creativity, and intentionality, which they view as essential qualities of true art. & ai, art, generated, crazy, real, fuck, hate, sad, bad, shit, program & 11.43\% \\
\cmidrule{1-5}
4 & Failure to Meet Expectations & This topic reflects users' disappointment when the AI-generated images fall short of their expectations. Criticisms often focus on perceived shortcomings in creativity, emotional expression, or technical execution, which fail to align with the anticipated artistic standard. & damage, hand, emotional, hear, remix, eyes, expecting, brain, think, forget, remember & 8.99\% \\
\cmidrule{1-5}
5 & Scary and Disturbing & This topic captures users' reactions to AI-generated images that are perceived as horrifying, unsettling, or irrational. These reactions often stem from unnatural features, eerie aesthetics, or a sense of the uncanny, evoking discomfort or fear. & dark, scary, souls, minecraft, wars, hand, terraria, style, fortnite, potter, art & 8.53\% \\
\cmidrule{1-5}
6 & Perceived Theft of Art & This topic reflects users' belief that AI-generated images are created by unfairly copying or appropriating the work of human artists. As a result, users feel a sense of betrayal or injustice, criticizing AI for undermining originality and ethical artistic practices. & did, use, make, theft, dirty, discord, program, wrong, cheating, steal, generate & 7.65\% \\
\cmidrule{1-5}
7 & Appalling Superiority & This topic reflects users' shock and discomfort at the realization that AI painting abilities have surpassed those of human painters. This raises concerns about the diminishing value of human creativity and the potential replacement of artists by machines. & hard, crazy, pic, insanely, incredibly, unbelievably, side, hit, monkey, doom, shit & 6.50\% \\
\toprule[1.2pt]
\end{tabular}%
}
\caption{Results of topic modeling. The results capture diverse and polarized reactions to AI-generated art, ranging from admiration for technical skill to concerns about authenticity, creativity, and ethical issues.}
\label{table:topic_modeling}
\end{table*}

\subsection{RQ3 - First part: How are public sentiments toward AIGP?}

After comparing the comment sentiment scores of both video types, we found in Table \ref{table: sentiment} that human painting videos have significantly more positive comments than AIGP videos. It is undeniable that the number of positive comments about both types of paintings is higher than the number of negative comments, and this data suggests, to a certain extent, that many people have already accepted that AI painting exists. Some AI-created artwork is also well-liked. However, compared to human artworks, some people tend to harbor more negative emotions when discussing AIGP content on social media. Overall, people still prefer to see human artworks on social media platforms rather than AI-generated ones.

\subsection{RQ3 - Second part: What motivates people’s negative attitudes toward AIGP?}
To answer this question, we conducted a topic modeling analysis. The topic modeling revealed 14 common topics within the dataset, among which the top 7 topics with the highest proportion are listed in Table \ref{table:topic_modeling}. To avoid the influence of positive comments, all topics were extracted from 19,071 negative comments about AIGP posted on TikTok. 

From Table~\ref{table:topic_modeling}, \textit{Topic 1} indicates that people often express admiration for the technical detail, precision, and stylistic fidelity that make such works comparable to traditional paintings. However, this appreciation is frequently accompanied by discomfort or unease, with users describing the images as “too real” or disturbingly lifelike. Keywords such as like, real, style, and sick underscore both the praise and uneasiness, suggesting a paradoxical emotional response to hyperrealism.

\textit{Topic 2} reveals that users' initial admiration is often followed by critical reconsideration upon learning the image was AI-generated. The emotional shift reveals tension between aesthetic appreciation and concerns about authenticity and authorship. Keywords like damn, hate, shame, insane, and interesting highlight the conflicted sentiments. Users often question whether AI creations can be considered true art, despite acknowledging their technical brilliance.

\textit{Topic 3} shows that users explicitly reject AI-generated content as legitimate art. The theme emphasizes a strong emotional response, characterized by terms like crazy, real, bad, and program. Critics argue that such works lack essential artistic qualities like creativity, intention, and emotional depth. The use of emotionally charged and profane language suggests a deeply rooted resistance to accepting AI within traditional artistic frameworks.

\textit{Topic 4} showcases users’ disappointment with AI art that does not align with their expectations. The criticism centers on perceived deficiencies in emotional expression, conceptual depth, or creative innovation. Common terms such as damage, hand, eyes, and expecting suggest unmet artistic standards, particularly in rendering human features or conveying emotional nuance.

\textit{Topic 5} reveals that AI art is perceived as eerie, uncanny, or frightening. Users describe images using keywords like dark, souls, and horror, indicating discomfort with unnatural or surreal visual elements. This aligns with theories of the "uncanny valley," where hyperrealistic but imperfect representations evoke fear or cognitive dissonance.

\textit{Topic 6} expresses ethical concerns about AI's use of human-created work as training data. Users feel that AIGP is derived unfairly, amounting to intellectual theft. Words such as theft, wrong, stealing, and used reflect feelings of betrayal and frustration, with commenters calling out AI systems for appropriating creative labor without acknowledgment or consent.

\textit{Topic 7} highlights existential fears about AI’s capabilities surpassing those of human artists. Users express alarm at the speed and skill with which machines can now produce high-quality artworks. Keywords like unbelievably, side, monkey, and doom suggest anxiety over the implications for human creativity and employment, as well as concerns about the diminishing role of human agency in the creative process.

Each topic will be elaborated on with quotes in the following paragraphs.

\subsubsection{Hyperrealism as both admiration and discomfort}

After analyzing the topics, we found a slight overlap between \textit{Topic 1} and \textit{Topic 7}, and we discussed these two topics together. These topics reflected that the reason for people’s negative attitude toward AIGP is that AIGP has become too realistic and is perceived likely to affect our lives in the future. Indeed, the advancement of AI technology has become increasingly unpredictable, and we are facing a dilemma in that it is difficult to distinguish between human paintings and AIGP. This dilemma will become increasingly challenging to break free from, along with the development of AI technology. One controversial incident involved using a deepfake voice in the documentary film “Roadrunner” about celebrity chef Anthony Bourdain~\cite{gamage2022deepfakes}. The film incorporated an artificial voice to simulate Bourdain speaking, which he never originated or consented to. This means that we may not even know when others have privately used our voices and likenesses. So when users are browsing these AIGP, it's hard for them not to conjure up Paintings of the day when we can no longer tell which ones are AI-generated and which ones are humans' own. Examples are given below:
\begin{quote}
    ``\textit{If you don't say AI, you won't know AI.}''
    
    ``\textit{I am afraid that the unique feeling of AI will be eliminated by refining some of the real ones.}''
    
    ``\textit{Finally, it's getting harder for AI to discriminate...}''
\end{quote}
They all express deep concern about the development of AI and the future of deepfake. This is, therefore, one of the most important reasons for users to oppose AI, accounting for 12.82 percent.

\subsubsection{From Awe to Doubt: The Ambivalence Toward AIGP}

\textit{Topic 2} expresses the idea that when many users see AIGP, they are indeed attracted by the fine character portrayal, rich color palette, and other artistic features, but when they see the AI tags underneath the video, they start to hesitate and sway from side to side or even feel disgusted and ashamed of the paintings, like these typical comments:

\begin{quote}
    ``\textit{This looks so gorgeous! Shame it's AI Art tho}''
    
    ``\textit{Disappointed when I realised it's AI art, stop taking from the community and faking art}''
    
    "\textit{the way my smile went away when i noticed its AI}"
\end{quote}
Users have already recognized the painting and think it is really good artwork, but considering the AI technology behind the artwork, they turn from a positive to a negative attitude. Therefore, the keywords ``cute'' and ``interesting'' in this topic do not represent a positive emotion but mockery and self-congratulation.

\subsubsection{Rejection of AIGP as True Art}

\textit{Topic 3} illustrates that some users have a cognitive bias towards AI art, a tendency or preference based solely on subjective perceptions and thought patterns and that other factors do not influence this bias. This cognitive bias has also been confirmed in previous research on AI paintings~\cite{ragot2020ai}. Such a phenomenon is similarly confirmed in this paper. In their perception, AIGP can never be called art, as this comment says:

\begin{quote}
    ``\textit{Sadly AI art isn't art, this is just slide after slide of dogshit}'' 
    
    ``\textit{if it's made by AI it's not art}''
\end{quote}
Thus, the negative attitude of this category towards AIGP is entirely unjustified, at least on social media, where they do not indicate the reason for their distaste for AIGP. Furthermore, this unjustified non-acceptance may not be limited to reviews on AIGP but on all things related to AI, such as:

\begin{quote}
    ``\textit{I hate ai soo much}'' 
    
    ``\textit{The fact that its AI generated makes it boring and empty}''
\end{quote}

\subsubsection{AIGP Falls Short of Creative Expectations}

Many AIGP are generated based on real people, anime characters, or classic scenes. The concept of \textit{Topic 4} is that many people think that AI-generated characters in Paintings don't live up to their expectations or don't capture some of the character's traits, which leads to their negative attitudes about AIGP. Some comments are given below:

\begin{quote}
    ``\textit{i think AI did kyo n yuki dirty nooo}'' 
    
    ``\textit{AI inconsistency when it comes to skin colour}'' 
\end{quote}
These comments highlight the conflict and mismatch between the AI-generated characters and the Painting of the character in people's minds. Some avid fans of anime characters consider that AI ruined their favorite characters, and as a result, they show strong resentment and anger in the comments.

\subsubsection{The Uncanny and the Disturbing Nature of AI Imagery}

There are two meanings we can get from \textit{Topic 5}. Firstly, the word ``dark'' indicates that there may be more dark and eerie types of AIGP on TikTok, and this style of drawing makes people feel uncomfortable. On the other hand, AI-generated technology has always had some complex technical difficulties, such as the drawing of human limbs and skin color:

\begin{quote}
    ``\textit{AI is pretty inconsistent when it comes to darker skin colors and the hair accuracy... shame but it has nothing to do with me}''
\end{quote}
Some of the keywords in this topic, like ``potter'' and ``hand'' indicate that some human-themed AIGP on social media have weird limbs or postures. These Paintings are likely to horrify and disturb users, leading to negative comments.

\subsubsection{AI as a Threat to Artistic Ownership and Ethics}
The reason for the negative attitude towards AIGP, as explained in \textit{Topic 6}, is that AIGP's creative process is perceived as a form of plagiarism, and therefore, people feel cheated. Some typical examples are as follows:

\begin{quote}
    ``\textit{idk if im impressed or scared like these ai art are STEALIN artists art omd}''
    
    ``\textit{many AIs that seek to draw the human, what they do is plagiarise the art of a list of authors, is just coming out.}'' 
    
    ``\textit{AI art is theft}''
\end{quote}
People think that AI can generate high-quality paintings by plagiarising the artists' paintings, which are not thought out by themselves, and therefore, AI paintings should not be accepted and spread.

\section{Discussion}
Humans and AI have become inextricably linked. Many examples and studies have shown that human-AI collaboration will lead to better creativity in many industries~\cite{maerten2023paintbrush}. However, according to some negative news, public feelings seem to be ignored, while generative AI is increasingly widely used for creative activities. Such discussions about public consciousness and awareness when living with AIGC are very critical, which ensure the competency of AI implemented responsibly and judiciously~\cite{sengupta2024public}. Therefore, the purpose of this study is to analyze users' discussion content and motivations toward AIGP on social media.

\subsection{Hesitation and Tentativeness in Interaction with AIGP}
According to Nadarzynski’s study on public responses to AI-led chatbot services, “AI hesitancy” was prevalent during users’ interactions with chatbots~\cite{nadarzynski2019acceptability}. 

In addition, according to the theory of the "uncanny valley," researchers have argued that people consistently report discomfort when reading posts produced by AI~\cite{radivojevic2024human}. Specifically, the study found that AI-generated text often triggers feelings of unease or aversion due to subtle deviations from authentic human communication, such as unnatural tone, overly formal structure, or lack of emotional nuance. This discomfort contributes to AI hesitancy and negatively impacts user engagement by reducing users’ willingness to interact with content perceived as AI-generated. Therefore, given the consistency between these findings and the context of the current research, the conclusions from previous studies are likely applicable here as well. By comparing the degree of user interaction between AIGP videos and human painting videos, we found that people still exhibit hesitation when responding to AIGP on social media, even though the interaction does not involve services or device usage. This low level of willingness to engage is reflected in the number of likes, favorites, and comments on AIGP-type videos.

Taking into account the possible technical limitations of generative AI, we controlled for the moderating variable of IAQ. However, the moderating role of IAQ was not found to be significant. These findings are consistent with previous research, which suggests that AI-generated artworks often face negative biases, particularly in terms of perceived creativity and authenticity\cite{millet2023defending,zhou2023eyes}. This further supports the notion that an attitude of mistrust or rejection toward generative AI technology itself may persist. Even as the quality and diversity of generative models continue to improve, this labeling effect—namely, the belief that human-made art is inherently more valuable—remains a major barrier to the acceptance of AI-generated content. Such biases may, to some extent, limit the broader acceptance and dissemination of AI-generated art.

By introducing \textit{Fear Theory} into AI-related research, several studies have acknowledged that emerging technologies are often perceived as potential threats due to the risks they may pose to individuals or society as a whole~\cite{osiceanu2015psychological,khasawneh2018technophobia}. In the context of AI, we are currently undergoing a transition toward cities and societies increasingly governed by artificial intelligences ~\cite{cugurullo2023fear}. However, the full risks associated with this transformation remain unclear~\cite{yigitcanlar2020can}. Building on these unresolved questions, this study contributes new empirical findings to the discourse on generative AI risks and public attitudes. We derived seven topics from user comments on AIGP to explore the underlying causes of negative sentiment. Among them, fear of AIGP was explicitly identified as a reason for such attitudes. Additionally, themes of ambivalence and perceptions of AI-generated images being “too real” were also found to be prominent drivers of negative responses.

\subsection{Influence of AIGP to Art}
There is no doubt that AIGP has had a significant impact on the art industry, as can be seen from the results of topic modeling, with \textit{Topic 1} and \textit{Topic 7} all highlighting the fact that today's AIGP are so similar to human paintings that we cannot distinguish some of the high-quality AIGP from human paintings. On December 6, 2023, a post made it to the top of  Weibo (one of China's most famous social media platforms, similar to Twitter) hotlist, which was along the lines of the mascot for the Spring Festival Gala announced by China Central Broadcasting Television (CCTV) being thought to have been generated by AI. Some users even posted long posts analyzing various details to prove that it was generated by AI. Although an official announcement was made to reply to the challenge, and the designer's original design was released, users still firmly believed it was AI-generated and accused the designer of this behavior. As a result, the emergence of AIGP has added many burdens to artists. As far as painters are concerned, they not only have to design creative and high-quality artworks but also need to worry whether their works will be labeled as AI and blamed. Some high-level painters have absolute confidence that their paintings can be better than AIGP. Still, they are afraid that the paintings they spent a lot of time and energy to create will be regarded as AI-generated. Therefore, this may be one of the reasons why AIGP is likely to cause many excellent artists to lose their creative drive and support anti-AI events.

\subsection{Implications for social media platform design}

This paper explored the effect of image aesthetic quality on user engagement, and the experimental results confirmed that the image aesthetic quality of paintings on TikTok does not have a significant impact on users' interactions. We found that users showed more robust emotional support and willingness to interact with some human paintings with relatively low aesthetic quality. This indicates that when browsing paintings on social media, users prefer to see the creator's continuous improvement process, even if the work is not perfect. However, in AI-generated paintings, we rarely see such emotional support. Therefore, this finding can provide some valuable insights into the study of social support in online communities, an important topic in social media research.

In addition, we analyzed and compared the sentiment of comments on AIGP and human paintings. The results showed that positive comments generally dominated people's comments on both types of paintings, but the rate of negative comments on AIGP videos was significantly higher than that on videos of human paintings. It means that there are still a lot of people who have a negative attitude towards AIGP, and many people are likely unwilling to see AI-generated Paintings on social media platforms. Therefore, in order to further explore what makes people hold negative attitudes toward AIGP content on TikTok, we conducted topic modeling of negative AIGP comments and finally identified seven main reasons. These reasons provide suggestions and guidance for optimizing future AI-generated models, from which we can analyze people's preferences to determine which AIGP to put on social media to attract users' attention and promote an intimate connection between the platform and users.

One of our findings is that people prefer viewing human-created paintings. This suggests that platforms can moderately reduce the promotion of AIGP and encourage artists to upload their own work, which can increase user engagement. On the other hand, we find that people's comments on AIGP videos express more negative emotions, which suggests that platforms may need to pay more attention to comments related to such videos to avoid vicious incidents.

\subsection{Limitations and Future Work}

This study uses user-generated content (UGC) on the TikTok platform as the source of experimental data, so data collection was done on a single social media platform. The TikTok platform provides a rich dataset and user voices that are relatively close to authentic expressions. However, it also presents certain limitations. Firstly, TikTok's user base tends to skew toward younger demographics~\cite{omar2020watch}, potentially limiting the generalizability of findings to broader populations. Secondly, we must acknowledge that the TikTok recommendation algorithms may introduce biases in the data. The algorithm’s non-random nature could skew results by over-representing certain user groups, such as tech enthusiasts for AI-generated art and traditional art enthusiasts for human-created art. Unlike randomized controlled trials, recommendation systems do not ensure balanced audience groups, which can lead to selection and allocation biases. Currently, there is no straightforward solution to address this issue, and it remains an open challenge. Future research will need to explore methods for mitigating these biases, perhaps by incorporating more controlled data collection strategies or by analyzing user characteristics within the dataset. Although we attempted to extend our survey to other mainstream platforms, many, like Twitter or Facebook, are not transparent about their data, making it challenging to gather comparable data. Consequently, the accuracy and generalizability of our results are constrained by this platform-specific limitation. To address this in future work, we plan to expand our data collection beyond TikTok to other platforms, such as Twitter, Facebook, and Reddit, and explore how user engagement and perception may vary across these different social media contexts.

Furthermore, while this study focused on comparing AI-generated and human-created art, the impact of the content of the paintings on audience feedback was not fully explored. In future work, we plan to further investigate how different characteristics of the artwork—such as style, complexity, and context—may influence user engagement and perceptions. By exploring these factors, we aim to provide a deeper understanding of how content attributes affect audience feedback, which could contribute to more nuanced comparisons between AI-generated and human-created art.

While the number of active users on TikTok is still the highest among all social media platforms, and all AIGP (AI-Generated Painting) videos must be tagged with "AI" (a feature not commonly found on other platforms), we did not focus solely on numerical data such as likes, shares, or views. We also analyzed tens of thousands of user comments, which provide deeper insights into the attitudes and emotions of users towards the two types of paintings. The integration of these various data points ensures the rigor of our findings. In future work, we plan to investigate these issues more deeply using alternative methods such as questionnaires or real-life interviews to validate our results and expand the scope of our research.

\section{Conclusion}

Because of TikTok's huge user groups and development speed far exceeding that of other platforms such as Facebook and Twitter, people's interactions and perceptions towards AIGP on the TikTok platform can reflect a social phenomenon to a certain extent. Therefore, this paper investigates the user engagement and emotional tendency of comments on AIGP on social media based on the TikTok platform and compares it with human drawings. Finally, it is concluded that people are more willing to interact with the content of human paintings on social media and have higher positivity towards their comments. We also considered the potential moderating role of the IAQ in human-AIGP interactions, and the results did not support this hypothesis. Finally, seven main themes derived from the topic modeling, including hyperrealistic qual- ity, ambivalent reactions, perceived theft of art, etc., indicated why some people still have negative perceptions towards AIGP.

\section{Disclosure statement}
No potential conflict of interest was reported by the author(s).

\section{Funding}

This research was supported by the 2021 CCF-Tencent Rhino-Bird Research Fund, the Research Matching Grant Scheme (RMGS), under Grants Number 9229095 ; and the 2024 CCF-Tencent Rhino-Bird Research Fund, the Research Matching Grant Scheme (RMGS) , under Grants Number 9229164.
\bibliography{reference}

\section{Author Biographies}
\subsection{Jiajun Wang}
Pre-Doctoral student and research assistant of School of Information Systems and Technology Management at University of New South Wales. Previously worked as a research assistant at City University of Hong Kong. His research interests include Information Systems and Human-Computer Interaction.

\subsection{Xiangzhe Yuan}
PhD student in Computer Science at the University of Iowa. Previously worked as a research assistant at City University of Hong Kong. Holds an MSc from Hong Kong Baptist University. Research focused on HCI and Human-AI Collaboration.

\subsection{Siying Hu}
Siying Hu is a Human-Computer Interaction researcher with a Ph.D. from the City University of Hong Kong. Her research lies at the intersection of AI for Science and Responsible AI, where she develops collaborative frameworks for material science discovery and is interested in the sociotechnical challenges of AI privacy and security.

\subsection{Zhicong Lu}
Assistant Professor of Computer Science at George Mason University. Research focuses on HCI, social computing, and generative AI for enhancing creativity, trust, and knowledge sharing. Previously at City University of Hong Kong. PhD from University of Toronto; MSc and BEng from Tsinghua University.

\end{document}